\documentstyle [12pt,epsf]{article}

\catcode`@=12
\newcommand{\be}{\begin{equation}}
\newcommand{\ee}{\end{equation}}
\newcommand{\bea}{\begin{eqnarray}}
\newcommand{\eea}{\end{eqnarray}}
\def\a{\alpha} \def\b{\beta} \def\f{n_f}
\def\d{n_d} \def\h{n_H}
\begin{document}
\thispagestyle{empty}
\begin{flushright} SINP/TNP/01-11\\UCRHEP-T307\\IC/2001/33\\May 2001
\end{flushright}
%\vspace{0.3in}
\begin{center}
{\Large \bf Supersymmetric Model of Neutrino Mass\\
and Leptogenesis with String-Scale Unification\\}
\vspace{0.7in}

{\bf Biswajoy Brahmachari$^{a}$, Ernest Ma$^{b}$, Utpal Sarkar$^{c,d}$\\}

\vskip 1cm

(a) Theoretical Physics Group, Saha Institute of Nuclear Physics,\\
AF/1 Bidhannagar, Calcutta 700064, INDIA \\
\vskip .5cm

(b) Physics Department, University of California,\\ 
Riverside, California, CA 92521, USA\\
\vskip .5cm

(c) Physics Department, Visva Bharati University,\\
Santiniketan 731235, INDIA\\
\vskip .5cm

(d) Abdus Salam International Centre for Theoretical Physics,\\
Strada Costiera 11, Trieste 34100, ITALY\\
 
\end{center}

\vskip 0.5in

\begin{center}
\underbar{Abstract} \\
\end{center}

Adjoint supermultiplets (1,3,0) and (8,1,0) modify the evolution 
of gauge couplings.  If the unification of gauge couplings occurs at the 
string scale, their masses are fixed at around $10^{13}$ GeV. 
This scale coincides with expected gaugino condensation scale in the
hidden sector $M_{string}^{2/3} m^{1/3}_{3/2} \sim 10^{13}$ GeV. 
We show how neutrino masses arise in this unified model which naturally
explain the present atmospheric and solar neutrino data.  The
out-of-equilibrium  decay of the superfield (1,3,0) at $10^{13}$ GeV may
also lead to a lepton  asymmetry which then gets converted into the
present observed baryon asymmetry of the Universe.

\newpage
\baselineskip 24pt

It is interesting to study the adjoint SU(2) triplet with no hypercharge 
$T\equiv (1,3,0)$ having a mass approximately $10^{13}$ GeV.  In a 
supersymmetric theory, the lepton superfield $L_i \equiv (\nu_i, l_i)$, the 
Higgs superfield $H_2 \equiv (h_2^+, h_2^0)$, and the superfield $T \equiv 
(T^+, T^0, T^-)$ may be connected by the Yukawa coupling $h^i_T~L_i~H_2~T$. 
As $H_2$ gets a nonzero vacuum expectation value (VEV) given by  $v_2 \equiv 
v~\sin \beta$, where $\tan \beta \equiv \langle h_2^0 \rangle/\langle h_1^0 
\rangle$, the neutrino $\nu$ pairs up with the fermion $T^0$ to form a 
Dirac mass and because $T$ has an allowed Majorana mass $M_2$, a seesaw 
mass is generated for $\nu$: \cite{ma98}
\be
m_\nu = {h^2_T~v^2~\sin^2~\beta \over 2 M_2},
\ee
where the extra factor of 2 comes from the fact that $T^0$ couples to 
$(\nu h_2^0 + l h_2^+)/\sqrt 2$. 
The difference between this and the canonical seesaw mechanism \cite{seesaw} 
is the use of an SU(2) triplet instead of a singlet.  This means that whereas 
the latter 
has negligible influence on the evolution of gauge couplings, the former 
changes it in a significant way.  It is thus possible to have gauge coupling 
unification at the string scale \cite{string} with $M_2 \sim 10^{13}$ GeV 
as well as a realistic theory of neutrino mass and leptogenesis consistent 
with present atmospheric and solar neutrino experiments \cite{atm,sol}, 
as shown below. 
   
One-loop string effects could lower the tree-level value of the 
string scale 
$M_{string}= g_{string} M_{Plank}$
somewhat, and one calculates \cite{stu} that the string
unification scale is modified to 
\be
M_{string}=g_{string} \times 5.27 \times 10^{17}~{\rm GeV} \simeq
5.27 \times 10^{17}~{\rm GeV}.
\ee
Furthermore, string models having a $ G \times G $ structure, when
broken to the diagonal subgroup, naturally contain adjoint scalars with
zero hypercharge. In this paper, we minimally extend the canonical 
supersymmetric standard model by including the superfields 
$T\equiv (1,3,0)$, $O\equiv (8,1,0)$ and $S\equiv (1,1,0)$ \cite{drtjones}.  
We will show that if the 
unification of gauge couplings occurs at the string scale, two-loop 
renormalization-group equations (RGE) will fix the masses of $T$ and $O$ 
at the well-motivated intermediate scale $M_I \sim M_{string}^{2/3} 
m^{1/3}_{3/2} \sim 10^{13}$ GeV, which turns out 
to be precisely the mass scale $M_2$ for $T \equiv (1,3,0)$ in Eq.~(1). 
In our RGE analysis, we consistently include the effects of all Yukawa 
couplings, among which are the constraints from our present knowledge of the 
neutrino mass matrix to account for the observed atmospheric and solar 
neutrino oscillations.  (In previous papers \cite{string}, this important new 
possibility was not recognized.) 

The dimensionless Yukawa couplings of this model in standard superfield notation is given 
by
\begin{equation}
{\cal L}= 
h_\tau~[L~H_1~\overline{E}]+~h_b~[Q~H_1~\overline{D}]
+h_t~[Q~H_2~\overline{U}]+h_T~[L~H_2~T]+h_S~[L~H_2~S]+h_\lambda~[T~T~S],
\end{equation}
where we have introduced a singlet $S \equiv (1,1,0)$, the utility of which 
will be explained later.
\begin{table}[htb]
\begin{center}
\[
\begin{array}{|c||c||c|}
\hline
Superfields& SU(3)_c \times SU(2)_L \times U(1)_Y & Anomalous~Dimension\\
\hline
L& (1,2,-{1 \over 2} \sqrt{ 3 \over 5}) & { 1 \over 16 \pi^2}~[ h^2_\tau 
+h^2_T+h^2_S- { 3 \over 2} g^2_2 - { 3 \over 10} g^2_Y ]\\ 
\overline{E}&(1,1,\sqrt{ 3 \over 5}) & { 1 \over 16 \pi^2 }~[2 h^2_\tau 
- { 6 \over 5} g^2_Y]  \\ 
\overline{D}& (\overline{3},1,{ 1 \over 3} \sqrt{ 3 \over 5}) & { 1 \over 
16 \pi^2}~[2 h^2_b-{ 8 \over 3} g^2_c - { 4 \over 30} g^2_Y] \\
\overline{U}&(\overline{3},1,-{2 \over 3} \sqrt{ 3 \over 5})& { 1 \over 16 
\pi^2}~[ 2 h^2_t-{ 8 \over 3} g^2_3 - { 8 \over 15} g^2_Y] \\
Q&(3,2,{ 1 \over 6} \sqrt{ 3 \over 5}) & { 1 \over 16 \pi^2}~[ h^2_t + 
h^2_b - { 8 \over 3} g^2_3 - { 3 \over 2} g ^2_2 - { 1 \over 30} g^2_Y]\\
H_1&(1,2,-{ 1 \over 2} \sqrt{ 3 \over 5})& { 1 \over 16 \pi^2}~[h^2_\tau+ 
3 h^2_b - { 3 \over 2} g^2_2 - { 3 \over 10} g^2_Y]\\
H_2&(1,2,{ 1 \over 2} \sqrt{ 3 \over 5}) & { 1 \over 16 \pi^2}~[3 h^2_t 
+h^2_T+h^2_S - { 3 \over 2} g^2_2 - { 3 \over 10} g^2_Y]\\
\hline
T & (1,3,0) & { 1 \over 16 \pi^2}~[h^2_T+h^2_\lambda - {
4 } g^2_2 ]\\
S & (1,1,0) & { 1 \over 16 \pi^2}~[3~h^2_\lambda+2~h^2_S ]\\
O &(8,1,0) &{ 1 \over 16 \pi^2}~[- {6} g^2_3 ]\\
\hline
\end{array}
\]
\end{center}
\caption{Representations and anomalous dimensions of superfields}
\label{table2}
\end{table}

At the two-loop level, the evolution of the gauge couplings is 
governed by the following equation, where we have defined
$t = ln ~\mu / 2 \pi$.
\be
{d \alpha_i \over d t} = {b_i}
\alpha^2_i + \sum_{j} {b_{ij} \over 4 \pi } \a^2_i \a_j -\sum_{k}   
{a_{ik} \over 4 \pi } \a^2_i Y_k, \label{2lrg}
\ee
The one-loop coefficients $b_i$ and the two-loop coefficients $b_{ij}$
can be easily derived \cite{drtjones}. Also the effect of the Yukawa
couplings on the running of the gauge couplings is brought in by the
coefficients $a_{ij}$. They are given by
\begin{equation}
b_i= \left(
\begin{array}{c} \frac{33}{5} \\
1 +2~n_T \\ -3 +
3~n_O\end{array} \right),\,~~~ 
a_{ik}=
\left(\begin{array}{cccccc} \frac{26}{5}& 
\frac{14}{5} & \frac{18}{5} & \frac{6}{5} & \frac{6}{5} &0 \\ 
6 & 6 & 2 & 6 & 2 & 4 \\ 
4 & 4 & 0 &0 &0 &0
\end{array} \right),
\ee
and 
\be
b_{ij}= \left( \begin{array}{ccc}
\frac{199}{25} & \frac{27}{5} &
\frac{88}{5} \\ 
\frac{9}{5} & 25 + 24~n_T & 24 \\ 
\frac{11}{5} & 9 & 14+ 54~n_O
\end{array}
\right),
\ee
In the matrix $a_{ik}$ the index $k$ refers to 
$Y_t,Y_b,Y_\tau,Y_T,Y_S,Y_\lambda$.  In the evolution equations we have 
generically used the notations $Y=h^2 / 4 \pi$ and $\alpha=g^2 / 4 \pi$.

As we know, we must also run the Yukawa couplings which are involved in the 
running of the gauge couplings. The RGE for a typical trilinear Yukawa 
term $d_{abc}~\phi_a~\phi_b~\phi_c$ is \cite{drtjones} 
\begin{equation}
\mu { \partial  \over  \partial \mu} d_{abc} = \gamma^i_a d_{ibc} + 
\gamma^j_b  d_{ajc} + \gamma^k_c d_{abk}  \label{yrg}
\end{equation}  
We now apply Eq.~(7) to the Yukawa couplings of interest. We thus get the 
evolution equations for the extra Yukawa couplings $h_T,h_S,h_\lambda$ 
as well as their influence on the evolution of the other relevant 
Yukawa couplings. Here also we put $t = ln ~\mu / 2 \pi$.
\begin{eqnarray}
{\partial Y_t \over \partial t}
&=& [6 Y_t+Y_b+Y_T+Y_S-{16 \over 3} \alpha_3 -3\alpha_2-{ 13 \over 15}
\alpha_Y] 
Y_t, \\ 
{\partial Y_b \over \partial t}
&=& [Y_t+ 6 Y_b+Y_\tau - { 16 \over 3} \alpha_3 - 3 \alpha_2 - { 7 \over 
15} \alpha_Y] Y_b, \\ 
{ \partial Y_\tau \over \partial t}
&=& [3 Y_b+4 Y_\tau  +Y_T+Y_S - 3 \alpha_2 - { 9 \over 5} \alpha_Y] 
Y_\tau, \\
{ \partial Y_T \over \partial t}
&=& [3 Y_t +Y_\tau + 3 Y_T + 2 Y_S + Y_\lambda - 7 \alpha_2 - { 3 \over 5}
\alpha_Y]
Y_T, \\
{\partial Y_S \over \partial t}
&=& [3 Y_t +Y_\tau + 2 Y_T + 4 Y_S + 3 Y_\lambda  - 3 \alpha_2 - { 3 \over 5}
\alpha_Y]
Y_S, \\
{\partial Y_\lambda \over \partial t}
&=& [2 Y_T + 2 Y_S + 5 Y_\lambda  - 8 \alpha_2]
Y_\lambda,
\label{rgsm} 
\end{eqnarray}
To calculate the masses of (1,3,0) $\equiv M_2$ and (8,1,0) $\equiv M_3$, 
we adopt the following procedure. We assume that the unification is
happening at the scale $M_X=5.2 \times 10^{17}$ GeV \cite{stu} 
with the unified coupling of $\alpha_X$.  We then use the two-loop RGE to 
evolve the couplings down to $m_Z$.  In doing so, we must properly cross 
the thresholds $M_2$ and $M_3$.  Once we get the values of the couplings at 
$m_Z$, we can numerically solve the set of quantities $\alpha_X,M_2,M_3$ 
using as input
\begin{equation}
\alpha_3(M_Z)=0.11-0.125,~\alpha_{2L}(M_Z)=0.03322,~\alpha_{1Y}(M_Z)=0.01688.
\label{inputs}
\end{equation}
Note that the running also depends on the Yukawa couplings present in
our model. We must have the top quark mass at around $174$ GeV. 
We keep the top-quark Yukawa coupling at its infrared fixed point
$h^2_t(M_X)/4\pi=1$ which gives a correct value of the top quark mass. 
We then vary $h_b(M_X)=h_\tau(M_X)=h$, 
or equivalently $\tan \beta$ as well as $\alpha_s(m_Z)$. 
The results are given in Fig.~1. 
The quantum chromodynamic (QCD) coupling does not feel the influence of 
$\tan \beta$ because QCD never gets broken.  Hence the $M_3$ solution is 
quite insensitive to $\tan \beta$, but $M_2$ does depend mildly on 
$\tan \beta$. [The contribution of adjoint scalars in nonsupersymmetric 
SU(5) was used \cite{babuma} in a similar way to increase the unification 
mass.]
\begin{figure}[htb]
\begin{center}
\epsfxsize=11cm
\epsfysize=11cm
\mbox{\hskip 0in}\epsfbox{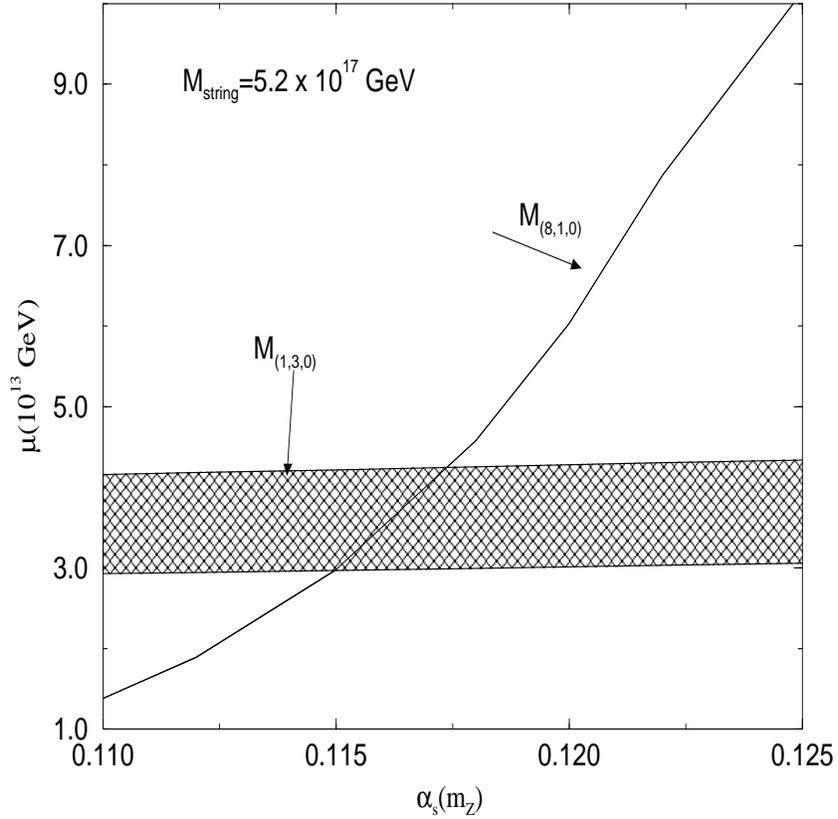}
\caption{These are the fitted values of $M_2$ and $M_3$. The flat curve
is $M_2$ which remains in the range $4 \times 10^{13}$ GeV. Note that
$M_3 \sim M_2$ for the central value of $\alpha_s=0.118$. The  
shaded region gives the effect of the extra Yukawa couplings.
$h_T(M_X),h_S(M_X), h_\lambda(M_X)$ takes all values between 0 and 3.54}
\label{fig1}
\end{center}   
\end{figure}

To first approximation, we let $T^0$ couple to $(\nu_\mu + \nu_\tau)/\sqrt 2$ 
and set $m_\nu = 0.05$ eV in Eq.~(1) to account for the atmospheric neutrino 
data.  For $M_2 \simeq 3.7 \times 10^{13}$ GeV, this implies that $h_T \sin 
\beta \simeq 0.35$.  To account for the solar neutrino data, we need another 
massive neutrino. As pointed out in Ref.[1], we have the option of choosing 
a heavy singlet $S \equiv (1,1,0)$ of mass $M_1$.  This has the virtue of not 
affecting the existing good convergence of the gauge couplings at the string 
scale because it does not have any one-loop contribution. 
With both $T$ and $S$, we also have the bonus of CP violation in a hybrid 
model \cite{hybrid} of leptogenesis, instead of using two triplets 
\cite{triplet}.

Let $S$ couple to $s\nu_e + c(\nu_\mu - \nu_\tau)/\sqrt 2 + \zeta (\nu_\mu + 
\nu_\tau)/\sqrt 2$, where $\zeta << 1$, then the neutrino mass eigenvalues 
are $m_2 = h_T^2 v_2^2/2 M_2$, $m_1 = h_S^2 v_2^2/M_1$, and 0, with 
eigenstates $(\nu_\mu + \nu_\tau)/\sqrt 2$, $s\nu_e + c(\nu_\mu - \nu_\tau)
/\sqrt 2$, and $c\nu_e - s(\nu_\mu - \nu_\tau)/\sqrt 2$ respectively.
We can write down the neutrino mass matrix as
\be
m_\nu=\pmatrix{ 
s^2~m_1 &  {(c+\zeta) ~s ~m_1 / \sqrt 2} & {(- c+\zeta)~ s ~m_1 /
\sqrt 2} \cr {(c+\zeta) ~s ~m_1 / \sqrt 2} & [m_2 + 
(c+\zeta)^2 m_1]/ 2 & [m_2 + (-c^2 + \zeta^2) m_1] / 2 
\cr {(- c+\zeta) ~s ~m_1 / \sqrt 2} & [m_2  + (-c^2 + \zeta^2) 
m_1]/ 2 & [m_2 + (-c+\zeta)^2 m_1] / 2 }.
\ee
Using the sample values $m_1 = 7 \times 10^{-3}$ eV, $m_2 = 5 \times 
10^{-2}$ eV, and $s = 0.6 (c = 0.8)$, this implies that $(\Delta m^2)_{atm} 
= 2.5 \times 10^{-3}$ eV$^2$, $\sin^2 2\theta_{atm} = 1$, and 
$(\Delta m^2)_{sol} = 4.9 \times 10^{-5}$ eV$^2$, $\sin^2 2\theta_{sol} = 
4 s^2 c^2 = 0.92$, in good agreement with data\cite{data}.

\def\a{\alpha} \def\b{\beta} \def\f{n_f}
\def\d{n_d} \def\h{n_H}

The couplings which are relevant for generating a lepton asymmetry
of the Universe\cite{leptogenesis,fuya} in this scenario are
contained in
\be
{\cal L} = h_{T}^i~ [L_i~ H_2~ T] + {1 \over 2} M_2~ [T~ T] 
+ h_{S}^i~ [L_i~ H_2~ S] + {1 \over 2} M_1~ [S~ S] 
+ h_\lambda ~[T~ T ~S],
\ee
where we have considered one triplet $T$ and one singlet $S$.  Note that 
with only one triplet or only one singlet, there cannot be any CP violation. 
The Majorana mass terms $M_2$ and $M_1$ for the triplet and singlet 
superfields violate lepton number and set the scale of lepton-number 
violation in this model. This scale has been determined by the evolution 
equations for the gauge couplings to be of the order $10^{13}$ GeV.  Note 
that all 3 new superfields $(S, T, O)$ are contained in the \underline {24} 
representation of SU(5), so it is not unreasonable for them to be at the 
same mass scale.

Their Yukawa couplings allow the triplet and singlet superfields to decay 
into final states of opposite lepton number.
\bea
T, ~S &\to& L + H_2 ~~{\rm and}~~ \bar L + \bar H_2.
\eea
There are one-loop vertex diagrams interfering with the tree-level
decay diagrams of $T$ and $S$, which will give rise to CP violation in 
these decays (see Fig.~2).  This CP asymmetry will then generate a lepton 
asymmetry of the Universe.  Unlike other models of leptogenesis where two or 
more heavy particles of the same type are used, there are no self-energy 
diagrams contributing to the CP asymmetry in this model.
\begin{figure}[htb]
\mbox{}
\vskip 1.25in\relax\noindent\hskip -.2in\relax
\includegraphics{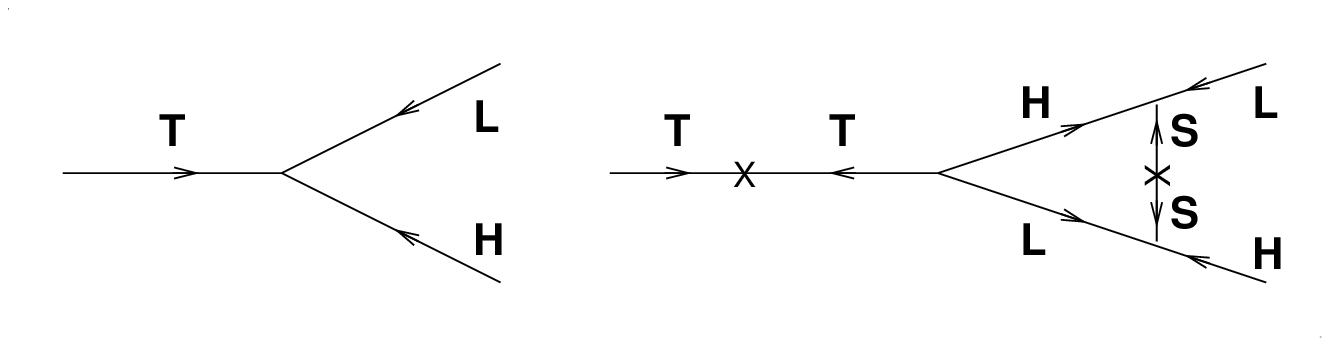}
\caption{Tree-level and one-loop vertex diagrams for the decay
of the triplet $T$ resulting in a lepton asymmetry.}
\label{fig2}
\end{figure}

Assuming $M_1 >> M_2$, the lepton asymmetry is generated by
the decay of the triplet superfield $T$. The singlet $S$ enters in
the loop diagram to give CP violation. The amount of asymmetry 
thus generated is given by
\be
\epsilon_L = {1 \over 8 \pi}  \left( {M_2 \over M_1} \right) \zeta^2 
{{\rm Im} [ (h_{S})^2 (h_{T}^{*})^2 ]  \over |h_{T}|^2},
\ee
where the factor $\zeta$ comes from the overlap between the neutrino states 
which couple to $T$ and $S$. 
In this scenario, the triplet superfield $T$ has gauge interactions,
which will bring its number density to equilibrium through the interaction 
$T + T \to W \to L + \bar L$.  However, the decay and the inverse decay of 
$T$ will be faster if we take $h^2_T(M_2)/4\pi >> \alpha_2(M_2)$. From our
RGE analysis we get $\alpha_2(M_2) \sim 1/25$. Since an asymmetry is only 
generated by a departure from equilibrium, 
interactions faster than expantion rate of the universe will bear 
an additional suppression factor in the asymmetry they generate. This factor 
can be estimated by numerically solving the full set of Boltzmann equations. 
We borrow the result from Ref.\cite{boltzman} that when $\Gamma / H$
is 5, the supression factor is 0.02 when it is 1000 the supression factor
can be as large as $8 \times 10^{-6}$. Here we make a rough estimate for
our case by taking the Yukawa interaction and neglecting the gauge
interaction and use an approximate supression factor,
\be
{\cal K} = \left( {H \over \Gamma} \right)_{T=M_2}, 
\ee
where $H = 1.7 \sqrt{g_*} ~T^2 /M_{Pl}$ (with $g_*$ the number of 
relativistic degrees of freedom) is the Hubble expansion parameter and 
$\Gamma = h_T^2 M_2/8\pi$ is the decay width of $T$. 
From our choice of numerical values for the neutrino mass matrix, we 
get an asymmetry 
\be
\epsilon_L = 3.4 \times 10^{-4} ~\zeta^2 \left(
{\sin 2 \delta \over \sin^2 \beta} \right) {\cal K},
\ee
where $\delta$ is the relative phase between $h_S$ and $h_T$. 
Using $M_2 = 3.7 \times 10^{13}$ GeV, $\zeta \sim 0.05$, and $\delta \sim 
0.01$, we then get a lepton asymmetry $\epsilon_L \sim 10^{-10}$ as required. 
A numerical solution of the Boltzmann equations can 
give errors introduced in parameters $\delta$ and $\zeta$ of this simple 
estimate. We plan to report this analysis in a future
publication. In this case, the
amount 
of lepton asymmetry is directly related to
the neutrino masses valid for atmospheric and neutrino oscillations as well 
as the intermediate scale required for string-scale unification. 
The scale of supersymmetry breaking in the hidden
sector (for a particular choice of the hidden sector fields) in this 
scenario may also be $10^{13}$ GeV, hence this particular intermediate 
mass scale allows us to have a consistent description of string-scale 
unification, neutrino mass, leptogenesis, as well as supersymmetry breaking.

If $M_2 >> M_1$, it will be the decay of the singlet $S$ which generates
the lepton asymmetry. In this case the singlet does not have any gauge
interactions but its Yukawa interaction will be similar to that of the 
triplet in the previous case.  Hence the amount of lepton asymmetry is 
again similar, except that the roles of $M_1$ and $M_2$ are reversed. 
Finally, this lepton asymmetry gets converted into the present baryon 
asymmetry of the Universe from the action of the $B + L$ violating 
electroweak sphalerons \cite{krs}, in analogy with the canonical 
leptogensis decay of heavy right-handed neutrinos \cite{fuya}.

Finally we would consider restrictions imposed
by inflationary scenarios as lepton asymmetry should be
created after the reheating starts after the inflation. 
For example if lepton number violation takes place at $10^{12}$
GeV and the upper bound on reheating temperature is $10^6$ GeV
it rules out the corresponding mechanism of leptogenesis. In our case the
mass scale $m_{3/2}$ is closely
related to $M_I$ and $m_{string}$ via
$M_I \sim M_{string}^{2/3}
m^{1/3}_{3/2}$. Furthermore the values used in the
RGE
analysis does not include threshold effects and ``smoothed''
threshold functions. The value of $M_I$ should be taken at best
as a guiding value. Taking $m_{3/2} \ge 5 (200) $ TeV we get
$M_I \ge 1.1 (3.8) \times 10^{13}$ GeV which is consistent with
value of the triplet mass obtained in Fig. \ref{fig1}.
Such a heavy gravitino should decay otherwise it will
over-close the universe. Now let us say that the gravitino decays 
predominantly to photon and photino. Upper bound on the
reheating temperature depends on the mass of
gravitino. We see from Figure (17) of 
the reference \cite{holtman} that for $m_{3/2}$ more than
5 TeV, reheating temperature upper-bound 
is more than $10^{13}$ GeV. Furthermore the produced photon
may further produce hadrons\cite{hadron}. In that case we get from
Figure (14) of reference \cite{kohri} that for $m_{3/2}$ more
than 200 TeV, the reheating upper-bound is more
than $10^{13}$ GeV. In both these cases our scenario
is consistent with post inflationary reheating. Infact note
that from RGE analysis we get values of $M_I$ which actually gives
$m_{3/2}$ in the 200 TeV range in a natural way. However the 
dominant decay mode of the gravitino may not
be photon and photino. The case where
the gravitino decays to a neutrino and sneutrino, when it is kinematically
allowed, has been studied
in \cite{moroi}. In this case neutrinos and sneutrinos produce
photons in cascade those interact and change predicted abundance of the
light elements which may differ from the observed values of
the abundance of light elements. In this case the upper
bound on the reheating temperature is tighter, which is around
$10^{12}$ GeV. This intermediate scale will produce a smaller
gravitino mass unless the string scale is lowered. In this
case the present scenario would be in trouble. Also, one must 
remember that there is experimental uncertainty in the determination of the 
abundance of light elements themselves such as the primordial fraction
of $^4$He\cite{izotov}. Finally in
various supersymmetric extensions of the standard model one can have light
axinos in the KeV range and gravitino decays to axino. In such scenarios
the upper bound on the reheating increases to $10^{15}$
GeV\cite{yanagida}.

A valid case can be made for larger soft masses of 100 TeV range
as it is good for suppressing supersymmetric flavor changing
neutran current and CP violation\cite{masiero} problems. In any 
case both supersymmetry and neutrino mass are physics
beyond the standard model. In our paper we have addressed neutrino
mass and a possibility of leptogenesis. In this model renormalization
group analysis has resulted an intermediate scale which gives   
$m_{3/2}$ in the range of 100 TeV. As long as there is no
fundamental reason why $m_{3/2}$ cannot be in the 100 TeV 
range our model stands correct.

In summary this is a scenario where the mass of the adjoint superfields 
$S \equiv (1,1,0), T\equiv(1,3,0)$ and $O\equiv(8,1,0)$ are all
approximately degenerate at $M_I \sim M_{string}^{2/3} 
m^{1/3}_{3/2} \sim 10^{13}$ GeV.
This scenario has been studied in the literature in the context
of string unification where it has been shown that
the unification of gauge couplings occur at $5.2 \times 10^{17}$ GeV.
In this paper we have shown that this scenario
can lead to a neutrino mass matrix which produces  
$(\Delta m^2)_{atm} = 2.5 \times 10^{-3}$ eV$^2$, 
$\sin^2 2\theta_{atm} = 1$, and $(\Delta m^2)_{sol} = 4.9 \times
10^{-5}$ eV$^2$, $\sin^2 2\theta_{sol} = 0.92$. This is in good
agreement with atmospheric and solar neutrino data. 
Furthermore, there are two ways that the triplet superfield $T$
may decay to $L$ and $H_2$, and in one of which the singlet $S$ 
resides in a loop. The interference of these decay amplitudes 
allows for the CP violation needed for leptogenesis. We have
shown that after we take into account the suppression 
in the generated lepton asymmetry due an approximate equilibrium
condition between the forward and inverse 
decays of $T$, the final lepton asymmetry emerges in the range
$\epsilon_L \sim 10^{-10}$ as required.

This work was supported in part by the U.~S.~Department of Energy under 
Grant No.~DE-FG03-94ER40837. BB thanks Probir Roy for communications on
gravitino decay modes.

\end{document}